\documentclass[reqno,12pt]{article}
\usepackage{amsmath,amsfonts,amssymb,amsthm,amstext,amscd,eucal}

\makeatletter \@addtoreset{equation}{section}

\makeatletter\renewcommand\section{\@startsection {section}{1}{\z@}%
                                   {-3.5ex \@plus -1ex \@minus -.2ex}%nn
                                   {2.3ex \@plus.2ex}%
                                   {\normalfont\large\bfseries}}
\renewcommand\subsection{\@startsection{subsection}{2}{\z@}%
                                    {-3.25ex\@plus -1ex \@minus -.2ex}%
                                   {1.5ex \@plus .2ex}%
                                  {\normalfont\bfseries}}

 \newcommand{\be}{\begin{equation}}
 \newcommand{\ee}{\end{equation}}
 \newcommand{\bea}{\begin{eqnarray}}
 \newcommand{\eea}{\end{eqnarray}}
 \newcommand{\nn}{\nonumber}
 
 \newcommand{\ri}{R^\mu_\nu}

\newcommand{\bse}{\begin{subequations}}
\newcommand{\ese}{\end{subequations}}
\begin{document}
\title{Holographic anomaly in 3d $f({\rm Ric})$ gravity}
\author{ Farhang Loran\thanks{e-mail:
loran@cc.iut.ac.ir}\\ \\
  {\it Department of  Physics, Isfahan University of Technology,}\\
{\it Isfahan, 84156-83111, Iran}}
\date{}
  \maketitle
 \begin{abstract}
   By applying the holographic renormalization method to the metric formalism of $f({\rm Ric})$ gravity
   in three dimensions, we obtain  the Brown-York boundary stress-tensor for backgrounds which asymptote to
   the locally AdS$_3$ solution of Einstein gravity.  The logarithmic divergence of the on-shell action
   can be subtracted by a non-covariant cut-off independent term which exchanges  the trace anomaly for a
   gravitational anomaly. We show that the central charge can be determined by means of BTZ holography or in terms of the Hawking effect of a Schwarzschild black hole placed on the boundary.
 \end{abstract}

\newpage
\tableofcontents

 %\keywords{Black Holes in String Theory, Supergravity Models }

%----------------------------------------------------------------------------------------------------------------

%\textcolor[rgb]{0.00,0.25,0.50}{\fixme{GH-term for Gauss-Bonnet 0910.5159}}

%\textcolor[rgb]{0.00,1.00,0.00}{\fixme{In 0910.5159, the ct is found in a perturbative approach in which my idea is followed. similarly in 9911152, while in 0103187 a hamiltonian approach is studied}}

%\textcolor[rgb]{0.30,0.60,0.20}{\fixme{Insist on Einstein solutions of f(Ric) [1301.1347] sect.4.If so then, h=0 by definition}}

%\textcolor[rgb]{1.00,0.00,0.00}{\fixme{Gaussian normal coordinate }}

%\textcolor[rgb]{1.00,0.00,0.00}{\fixme{show that h=0, specially note the critical point: [1301.1347] eq.(33), (34) }}

%\textcolor[rgb]{1.00,0.50,0.00}{\fixme{Asym. locally AdS vs. Asym AdS [1301.1347] p6 footnote1}}

%%\textcolor[rgb]{1.00,0.00,0.50}{\fixme{see Hohm-Tonni- the published paper for the 12ed term}}

%\textcolor[rgb]{1.00,0.00,0.50}{\fixme{Discuss dependence of the counter-term on the boundary condition [1006.3349][Cunliff] and [1106.4609]}}

\section{Introduction}
 Black hole physics is the essential ingredient of any quantum theory of gravity. In the context of AdS$_3$/CFT$_2$ correspondence, the CFT partition function of a BTZ black hole \cite{BTZ1,BTZ2} can be identified via a modular transformation in terms of the free energy of the vacuum which corresponds  to the thermal AdS$_3$ \cite{Maldacena:1998bw}, and the Cardy formula \cite{Cardy} reproduces the Bekenstein-Hawking black hole entropy \cite{Strominger-97}. The Virasoro algebra of the dual CFT is initially identified as the asymptotic symmetry algebra of the AdS$_3$ spacetime \cite{Brown-Henneaux}. For Einstein gravity, the corresponding central charge can be determined in terms of the holomorphic Weyl anomaly  \cite{Henningson-Skenderis}. In \cite{deHaro, Bala} the holographic stress-energy tensor is identified in terms of the Brown-York tensor \cite{Brown-York}.

 Inspired by the Brown-Henneaux approach to  the AdS/CFT correspondence \cite{Brown-Henneaux},
 it is natural to seek the extension of the duality to higher-derivative gravity in AdS$_3$.
 Since in three dimensions, the Riemann tensor can be given in terms of the metric ${\cal G}_{\mu\nu}$ and
 the Ricci tensor $\ri$,  $f(\ri)$  gravity, in which $f$ is a polynomial in $\ri$, is quite interesting. Massive gravity studied in \cite{NMG} is an example of such models.

 The first step towards holography is  identifying the CFT stress-tensor. Following \cite{witten}, the AdS/CFT correspondence implies that the expectation value of the stress-energy tensor of the dual   CFT can be identified with the Brown-York tensor \cite{Bala}. In order to obtain the Brown-York tensor,  one needs to identify the surface terms which are needed to make the  action stationary given only $\delta {\cal G}_{\mu\nu} = 0$ on the boundary.   For the two-derivative Einstein-Hilbert action, the surface term is the Gibbons-Hawking term \cite{Gibbons-Hawking}.

 For $f(R)$ models, in which $R$ denotes the Ricci scalar, one needs to cancel surface terms that depend on $\delta R$. In \cite{Madsen-Barrow}, the authors argue  that no such boundary terms exist in general.
 It is known that $f(R)$ gravity is  equivalent to Einstein gravity coupled to a scalar field. Of~course this equivalence relies on a conformal transformation  which can be in general   singular \cite{Hawking-Luttrell}. More precisely, $f(R)$ model in metric formalism is equivalent to $\omega=0$  Brans-Dicke theory \cite{Whitt}. From this point of view, $R$ carries the scalar degree of freedom and $\psi\equiv f'(R)$ is christened scalaron \cite{Frolov}. So it is reasonable to set $\delta R = 0$ on the boundary   \cite{Dyer-Hinterbichler}.  In the GR limit $f(R)\to R$ the scalar field decouples from the theory \cite{Olmo} and consequently there is no need to make any assumption on $\left.\delta R\right|_B$ in GR.\footnote{ In \cite{Madsen-Barrow}, it is shown that  the assumption $\left.\delta R\right|_B=0$ can be relaxed if the spacetime is assumed to be maximally symmetric i.e. assuming ${\tilde K}^{\mu\nu}\delta{\tilde K}_{\mu\nu}|_B=0$ and  $\left.\delta(n^\mu\nabla_\mu K)\right|_B=0$ where ${\tilde K}_{\mu\nu}$ is the traceless part of the extrinsic curvature $K_{\mu\nu}$, $n^\mu$ is the unit normal to the boundary, and $\nabla_\mu$ denotes the covariant derivative with respect to the Levi-Civita connection corresponding to ${\cal G}_{\mu\nu}$.}

  In the more general case of $f(\ri)$ gravity, different approaches  have been
  considered. For example,  in \cite{Myers} the surface terms are determined for general Euler density actions;
  in  \cite{Fukuma} this terms are given in a first order formulation of the theory, and in \cite{Cremonini},
  the surface terms are obtained in an on-shell perturbative  approach, i.e. one considers the higher derivative terms as perturbations to the Einstein-Hilbert action, and uses the field equations to compute the necessary boundary term.

  In order to find the Brown-York tensor, one also needs to determine the counter-terms
  which holographically renormalize  the action, i.e. make the action finite for asymptotically locally AdS
  backgrounds. For Einstein gravity, these terms are computed in \cite{Henningson-Skenderis,deHaro,Bala}. In \cite{Nojiri, Serg}, this method is generalized to $R^2$ models  and in \cite{Hohm-Tonni}, the corresponding counter-terms are obtained in the second order formulation involving an auxiliary tensor field. We intend to generalize these results to arbitrary $f(\ri)$ models in three dimensions.

  Actually, the Ostrogradski's theorem implies that $f(\ri)$ models are in general instable \cite{Woodard}.
  This instability is   explicitly shown e.g. in \cite{Waldram}, and is extensively studied in the case of massive gravity \cite{NMG}. We are not going to study the stability of $f(\ri)$ models here. Our goal is to obtain the holographically renormalized  Brown-York tensor for $f(\ri)$ gravity in backgrounds which asymptote to locally AdS$_3$ solution of  Einstein gravity,
    \be
 R_{\mu\nu}=-2\,\ell^{-2}{\cal G}_{\mu\nu}.
 \label{int-1}
 \ee
 In principle, if AdS/CFT correspondence can be generalized to higher-derivative gravity, then the instability
 of the $f(\ri)$ model can be realized in the dual CFT. So, in principle, the issue of stability could deepen our  understanding of  holography.

 The central charge of the dual CFT can be identified in terms of the Weyl anomaly \cite{Henningson-Skenderis}. In \cite{Imbimbo} a universal formula for the so-called  type A anomalies is obtained for $f(R)$ gravity. In particular, in three dimensions, the value of the central charge computed  by this method equals the value obtained  in \cite{Kraus-Larsen, Saida-Soda} which generalizes the results of \cite{Henningson-Skenderis} to higher-derivative models of gravity.
 By using these methods, one can determine the central charge without necessarily obtaining the stress-tensor.
 The central charge appears to be given by the Brown-Henneaux formula,  in which,  the Newton's constant $ G$ is screened by $\Omega$  defined by \cite{Hohm-Tonni,Kraus-Larsen,Saida-Soda},
 \be
  \left.f^\nu_\mu \right|_B=\Omega\,\delta^{\nu}_\mu,\hspace{1cm}f^\nu_\mu= \frac{d\, f}{d R^\mu_\nu}.
  \label{omega}
  \ee

  In this paper, we apply the holographic renormalization method to the $f(\ri)$ model in backgrounds that asymptote to locally AdS$_3$ spacetimes \eqref{int-1}. In the second-order formulation given by the action \cite{Waldram},
   \be
 S_{\rm{2nd}}= \int_V
 \left[f-f^\nu_\mu\left(\chi^\mu_\nu-R^\mu_\nu\right)\right],
 \label{second-order-action}
 \ee
 in which, $\int_V$ stands for $\int  d^{d+1}x \sqrt{{\cal G}}$ and $\chi^\mu_\nu$ is an auxiliary tensor field,
 one can simply follow the  method of \cite{Hohm-Tonni}.  In this formulation, $\delta\chi^\mu_\nu$ is assumed to be
 vanishing on the boundary, and the method of \cite{Henningson-Skenderis,deHaro} can be used, where, effectively, the Gibbons-Hawking term is given in terms of the screened Newton's constant.

  The higher-derivative formulation of the $f(\ri)$ model is given by the
  action,
  \be
  S=\int_V f(\ri).
  \label{higher-d-action}
  \ee
  In this case,  one needs to add a counter-term to compensate for the $\delta R$-dependent surface terms.
  As we discuss in section \ref{sec-ALAdS}, such a boundary  term is accessible in asymptotically locally AdS$_3$
  backgrounds, where, the traditional Fefferman-Graham expansion \cite{Fefferman-Graham-paper} is  available. We show that the resulting stress-tensor is essentially equivalent to the one obtained in the second-order formulation.

  We then turn to the on-shell value of the action, which, following \cite{witten} is an essential ingredient
  of holography, as it gives the leading term in the CFT partition function. It is known that there is a logarithmic
  divergence in the on-shell value of the action, which, can be subtracted by a cut-off dependent covariant
  counter-term \cite{Henningson-Skenderis,deHaro}. Here, we examine a cut-off independent term which appears to be not covariant. After adding this term, the trace anomaly disappears and a gravitational anomaly materializes instead. It is known that in two-dimensions,  gravitational anomaly and trace anomaly can be switched by adding a local counter-term \cite{David}. Here, we show that the value of the central charge can be determined in terms of the gravitational anomaly, by means of the holography of BTZ black holes or in terms of the Hawking effect of a Schwarzschild black hole placed on the boundary.

 The organization of the paper is as follows.  In section \ref{Sec-PBH}, following \cite{Imbimbo} we compute the Weyl anomaly in $f(\ri)$ model by studying bulk diffeomorphisms  corresponding to the Weyl transformation of the boundary metric. In section \ref{Sec-Einstein},
 we review the holographic renormalization  in Einstein gravity \cite{Henningson-Skenderis,deHaro}, and
 extend it to $f(\ri)$ gravity in section \ref{sec-Ricci-tensor}.  In section \ref{Sec-GA}, we study the gravitational
 anomaly that  appears when the logarithmic divergence is subtracted by means of a cut-off independent counter-term.
 Section \ref{Sec-discussion} is devoted to a short discussion about the CFT dual to $f(\ri)$ gravity.
 Some technical details are relegated to  appendices.
 %================================================================================================
 \section{Weyl anomaly in $f(\ri)$-model}\label{Sec-PBH}
 Assume a general gravitational action,
  \be
  S=\int_V f(\ri).
  \label{action0}
  \ee
    We are considering $f(\ri)$ as a function of $R^\mu_\nu={\cal G}^{\mu\rho}R_{\rho\nu}$, with all contractions made between raised and lowered indices so that the metric does not enter explicitly \cite{Waldram}. Under a bulk diffeomorphism, this action is invariant up to a boundary term
    \cite{Schwimmer-Theisen},
  \be
  \delta_\xi S=\int d^{d+1}x\, \partial_\alpha\left[\sqrt{{\cal G}} f(\ri)\xi^\alpha\right]=-\int_B\, n_\alpha\, \xi^\alpha f(\ri).
  \label{diffeo}
  \ee
  in which, $\int_B$ stands for $\int d^dx\sqrt{\gamma}$, where, $\gamma$ is the induced metric on the boundary and $n_\alpha$ is the {\em inward} pointing unit normal to the boundary.
  For an asymptotically locally AdS solution ${\cal G}_{\mu\nu}=\bar {\cal G}_{\mu\nu}$, the Weyl anomaly is given by this boundary term for a PBH (Penrose-Brown-Henneaux) transformation \cite{Imbimbo,Schwimmer-Theisen}. Details of this transformation is not important for us. What we are going to show is that, the Weyl anomaly of $f(\ri)$ model for an asymptotically locally AdS solution ${\cal G}_{\mu\nu}=\bar {\cal G}_{\mu\nu}$, equals the Weyl anomaly of the Einstein-Hilbert action with a cosmological constant term corresponding to the AdS background ${\cal G}^{\rm AdS}$ describing the asymptotic geometry of  $\bar {\cal G}_{\mu\nu}$, and a screened Newton's constant $G/\Omega$. To see this, one needs to compute the Taylor expansion of $f(\ri)$ around $\bar R_{\mu\nu}$, the Ricci tensor corresponding to $\bar {\cal G}_{\mu\nu}$,
  \be
  f(\ri)=f(\bar R^\mu_\nu)+\frac{d\,f}{dR^\mu_\nu} (R^\mu_\nu-{\bar R}^\mu_\nu)+{\mathcal O}(\ri-\bar R^\mu_\nu)^2.
  \label{Taylor}
  \ee
 Thus,
  \be
  \left.\delta_\xi S\right|_{{\cal G}=\bar {\cal G}}=-\left[{\Omega}\int_B n.\xi (R-2\Lambda)\right]_{{\cal G}={\cal G}^{\rm AdS}},
  \label{eq-deltaS}
  \ee
  in which, $\Omega$ is given by Eq.\eqref{omega}, and,
  \be
  2\Lambda=\left[ R-\Omega^{-1}f( R^\mu_\nu)\right]_{{\cal G}={\cal G}^{\rm AdS}}.
  \label{cosmol-cons}
  \ee
  In other words,
  \be
  \left.\delta_\xi S\right|_{{\cal G}=\bar {\cal G}}=\left.\delta_\xi S_{\rm EH}\right|_{{\cal G}={\cal G}^{\rm AdS}},
  \ee
  where,
  \be
  S_{\rm EH}=\frac{\Omega}{16\pi G}\int_V (R-2\Lambda).
  \label{Gs}
  \ee
  This result confirms that the Weyl anomaly in $f(\ri)$ gravity on asymptotically locally AdS backgrounds is given by  the Brown-Henneaux formula \cite{Brown-Henneaux} with a screened Newton's constant \cite{Kraus-Larsen}.
 %============================================================================================
 \section{Holographic renormalization in pure Einstein gravity}\label{Sec-Einstein}
 In this section, we review the holographic renormalization of Einstein gravity in asymptotically locally AdS$_3$
 spacetimes  \cite{Henningson-Skenderis,deHaro}.

  The AdS$_3$ solution of the Einstein field equation with a negative cosmological constant $\Lambda=-\ell^{-2}$,
 \be
 \Pi_{\mu\nu}=R_{\mu\nu}-\frac{1}{2}R g_{\mu\nu}+\Lambda g_{\mu\nu}=0,\hspace{1cm}\mu,\nu=0,1,2.
 \label{Sk-einstein}
 \ee
 is given by,
 \be
 ds^2=\frac{\ell^2dr^2}{4r^2}+r^{-1}(-dt^2+d\phi^2),
 \label{Sk-metric}\ee
 in which, $t=\ell^{-1}t_{\rm AdS}$. An asymptotically locally AdS solution in  {\em normal} coordinates is given by,
 \be
 ds^2=\frac{\ell^2dr^2}{4r^2}+\gamma_{ij}\,dx^i\,dx^j,\hspace{1cm}i,j=1,2.
 \label{Sk-asym-metric}
 \ee
 where, using  the {\em traditional} Fefferman-Graham asymptotic expansion \cite{Fefferman-Graham-paper},
 \be
 \gamma_{ij}=r^{-1}g_{ij}=r^{-1}g^{(0)}_{ij}+g^{(2)}_{ij}+h^{(2)}_{ij}\ln r+{\cal O}(r).
 \label{Sk-Fefferman}
 \ee
 In these coordinates, the boundary is located at $r=0$.   The extrinsic curvature of the boundary is given by,
 \be
 K_{\mu\nu}=\nabla_{\mu}n_\nu,
 \ee
 in which, $\nabla_\mu$ denotes the covariant derivative with respect to the Levi-Civita connection corresponding to the metric \eqref{Sk-asym-metric}, and,
 \be
 n^{\mu}=\left(\frac{2r}{\ell}, 0,0\right),
 \label{Sk-normal-vector}
 \ee
 is the {\em inward} pointing surface-forming normal vector.    The components of the extrinsic curvature are,
 \be
 K_{r\mu}=0,\hspace{1cm}K_{ij}=\frac{r}{\ell}\gamma_{ij,r}.
 \ee
 Eq.\eqref{Sk-einstein} implies that \cite{deHaro},
 \bea
 \label{Sk-h2}
 h^{(2)}_{ij}&=&0,\\
 \label{Sk-div-of t}
 {\cal D}^it_{ij}&=&0,\\
 \label{Sk-R-g2}
 {\cal R}^{(0)}&=&-2\ell^{-2}{\rm tr}g^{(2)},
  \eea
 where, $\gamma_{ij}$ and its inverse are used to lower and raise the Latin indices, while the trace operator `$\rm tr$' is defined
 in terms of $g^{(0)}_{ij}$. The covariant derivative ${\cal D}_i$ is defined with respect to the Levi-Civita connection
 corresponding to $\gamma_{ij}$,
  \be
  {\cal D}_i={\cal D}^{(0)}_i+{\cal O}(r),
  \ee
  in which, ${\cal D}^{(0)}_i$ is defined with respect to  $g^{(0)}_{ij}$, and  ${\cal R}^{(0)}$ is the corresponding scalar curvature. Finally,
    \bea
 t_{ij}&=&K_{ij}-(K+\ell^{-1})\,\gamma_{ij}\nn\\&=&\ell^{-1}\left(g^{(2)}_{ij}-g^{(0)}_{ij}{\rm
 tr}g^{(2)}\right)+{\cal O}(r).
 \label{Sk-t-ij}
 \eea
  For example, Eq.\eqref{Sk-div-of t} is given by the field equation
 $\Pi_{ri}=0$, which implies that,
 \be
 0=R_{ri}=\gamma^\nu_i[\nabla_\alpha,\nabla_\nu]n^\alpha={\cal
 D}_i\,t^i_j.
 \ee
 The last equality is obtained by noting that,
 \be
 {\cal D}_iK^i_j=\gamma^\mu_i\gamma^i_\nu\gamma^\rho_j\nabla_\mu K^\nu_\rho=(\delta^\mu_\nu-n^\mu n_\nu)\gamma^\rho_j\nabla_\mu K^\nu_\rho,
 \ee
 where, in order to obtain the first equality, we have used Lemma 10.2.1 in \cite{Waldbook}. The second equality is obtained by noting that $n.\nabla n_\nu=n_\mu K^\mu_\nu=0$.

 The Einstein-Hilbert action is given by,
 \be
 S_{\rm EH}=\frac{1}{2\kappa^2}\int_V(R-2\Lambda),
 \ee
 in which, $\kappa^2=8\pi\,G$.  The variation of the action with respect to  $\delta {\cal G}_{\mu\nu}$ is given by,
 \bea
 \delta S_{\rm EH}&=&\frac{1}{2\kappa^2}\int_V \left({\cal G}^{\mu\nu}\delta R_{\mu\nu}+\Pi_{\mu\nu}\delta {\cal G}^{\mu\nu}\right)\nn\\
  &=&\frac{1}{2\kappa^2}\int_B\left({\cal G}^{\mu\nu}\delta K_{\mu\nu}+\delta K\right)-\frac{1}{2\kappa^2}\int_V\Pi^{\mu\nu}\delta {\cal G}_{\mu\nu},
 \eea
 where, we have used Eqs.\eqref{10-11-12-m2}, \eqref{10-11-12-m3}, \eqref{deltaKij} and \eqref{deltaK}.
 The second term gives the Einstein field equation \eqref{Sk-einstein} and is vanishing on-shell. Henceforth, we drop this term. The first term depends on $n.\nabla\delta\gamma_{\mu\nu}$ and can be removed by adding the Gibbons-Hawking term,
 \be
 S_{\rm GH}=-\frac{1}{\kappa^2}\int_BK.
 \ee
 Thus,
 \be
 \delta S=-\frac{1}{2\kappa^2}\int_B(K_{\mu\nu}-K{\cal G}_{\mu\nu})\,\delta {\cal G}^{\mu\nu},
 \label{Sk-delta-S}
 \ee
  in which, $S= S_{\rm EH}+ S_{\rm GH}$.   The Brown-York tensor is defined by,
 \be
 T_{ij}=-\left.\frac{2}{\sqrt{\gamma}}\frac{\delta S}{\delta\gamma^{ij}}\right|_{\rm on-shell},
 \label{BY-definition}
 \ee
 where, $\delta\gamma_{\mu\nu}$ is the variation of the induced metric on the boundary, which obeys the constraint $n^\mu\delta\gamma_{\mu\nu}=0$.
 Furthermore, one assumes that $\delta n^\mu=0$.
 The minus sign in Eq.\eqref{BY-definition} reflects the fact that, one defines the energy-momentum tensor in terms of $\delta\gamma_{ij}$. Here, noting that $\delta\gamma^{ij}\sim{\cal O}(r)$ and $\sqrt{\gamma}\sim{\cal O}(r^{-1})$, we have given the Brown-York tensor in terms of $\delta\gamma^{ij}$. The idea is to identify $T_{ ij}$, after renormalization, with the expectation value of the stress-energy tensor of the dual
  CFT \cite{Bala},
 \be
 \left<{\cal T}_{ij} \right>_{\rm CFT}=T^{\rm ren}_{ij},
 \ee
 where, on the boundary, the indices are raised and lowered by $g^{(0)}_{ij}=r\left.\gamma_{ij}\right|_B$.
 Using Eq.\eqref{Sk-delta-S}, one obtains,
  \be
  \kappa^2\,T_{ij}=K_{ij}-K\gamma_{ij}=\ell^{-1}\gamma_{ij}+t_{ij}.
  \ee
  The first term is singular on the boundary and can be removed by adding a counter-term to the Gibbons-Hawking term \cite{Bala},
  \be
  S^{\rm reg}_{\rm GH}=-\frac{1}{\kappa^2}\int_B(K+\ell^{-1}).
  \label{GH-Eintein-counter-term}
  \ee
  Thus, the action is,
  \be
  2\kappa^2\, S=\int_V(R-2\Lambda)-2\int_B(K+\ell^{-1}),
  \ee
  and the regularized Brown-York tensor is,
  \be
  T^{\rm ren}_{ij}=\kappa^{-2}t_{ij}.
  \ee
 We still need to remove a logarithmic divergence in the on-shell value of the
 action \cite{deHaro}.  Recall that the on-shell value of the action gives
  the tree-level contribution to the free-energy of the boundary CFT \cite{witten}. Since,
  \be
 \sqrt{{\cal G}}=\frac{\ell}{2r}\sqrt{\gamma}=\frac{\ell \sqrt{g^{(0)}}}{2}\left(\frac{1}{r^2}+\frac{{\rm tr} g^{{(2)}}}{2r}+\cdots\right),
 \label{squart-G}
 \ee
 one verifies that,
 \be
 \int_V(R-2\Lambda)=-\lim_{\epsilon\to0}\frac{2}{\ell }\int d^2x \sqrt{g^{(0)}}\left(\frac{1}{\epsilon}-
 \frac{1}{2}{\rm tr} g^{{(2)}}\ln\epsilon\right)+{\rm finite}.
 \ee
 The regularized Gibbons-Hawking term $S^{\rm reg}_{\rm GH}$, removes the $\epsilon^{-1}$ term. Thus, using Eq.\eqref{Sk-R-g2}, one obtains \cite{Henningson-Skenderis,deHaro},
 \be
 2\kappa^2\,S_{\rm log-term}=-2\pi\ell\chi\lim_{\epsilon\to0}\ln\epsilon,
 \label{Sk-log}
 \ee
 in which, $\chi$ is the Euler-characteristic of the boundary,
 \be
 \chi=\frac{1}{4\pi}\int d^2x\sqrt{g^{(0)}}{\cal R}^{(0)}.
 \label{Euler}
 \ee
 Thus, the counter-term is a topological term and do not contribute
 to the Brown-York stress tensor \cite{deHaro}.

 Eq.\eqref{Sk-div-of t} implies that,
 \be
 {\cal D}^i T^{\rm ren}_{ij}=0,
 \ee
 and Eq.\eqref{Sk-R-g2} gives,
 \be
 {\rm tr}{T^{\rm ren}}=\frac{c}{24\pi}{\cal R}^{(0)},
 \ee
in which,  $c=3\ell/2G$ is the Brown-Henneaux central charge \cite{Brown-Henneaux}.
 It is important to note that the logarithmic divergence of the
 on-shell action \eqref{Sk-log} is given by the central charge
 \cite{deHaro,Kraus-Larsen},
 \be
 S_{\rm log-term}=-\frac{c\,\chi}{12}\lim_{\epsilon\to 0}\ln \epsilon.
 \label{Sk-chi}
 \ee
 %========================================================================================================================

%========================================================================================================================
 \section{Holographic renormalization of $f(\ri)$-model}\label{sec-Ricci-tensor}
 In the previous  section, we studied renormalization of the on-shell Einstein-Hilbert  action and the corresponding
 Brown-York stress tensor for asymptotically locally
 AdS$_3$ spacetimes. In this section, we study this problem in the  $f(\ri)$ model of
 gravity. In section \ref{sec-GB}, we discuss the generalization of
 the Gibbons-Hawking term in the second-order formulation and in the
 higher-derivative formulation of $f(\ri)$ gravity. In section \ref{sec-ALAdS}, we obtain the  surface
 terms for  asymptotically locally AdS$_{3}$  spacetimes, and study holographic renormalization of the  corresponding Brown-York
 tensor.
 \subsection{Surface terms}\label{sec-GB}
 The higher-derivative formulation of $f(\ri)$ gravity is given by the action \eqref{higher-d-action} which  is classically equivalent to a second  order action   given by Eq.\eqref{second-order-action}  \cite{Balcerzak}.
 The field equation for $\chi$ gives,
 \be
 \frac{d\, f^{\mu}_\nu}{d \chi^\alpha_{\beta}}(\chi^\nu_\mu-R^\nu_\mu)=0,
 \ee
 implying  that $\chi^\nu_\mu=R^\nu_\mu$ whenever  $\det \frac{d\, f^{\mu}_\nu}{d\, \chi^\alpha_{\beta}}\neq0$ \cite{Waldram}. It should be noted that this field
 equation does not depend on $\left.\delta\chi^\nu_\mu\right|_B$. As far as the auxiliary field is considered  as an independent
 field, one can assume that $\delta\chi^\mu_\nu$ is vanishing on the
 boundary \cite{Hohm-Tonni}.

 In the both formulations, one supplements the action with a boundary
 term,
 \be
 \int_B \left({\cal L}_{GH}+{\cal L}_{ct}\right),
 \ee
 in which, ${\cal L}_{GH}$ is the Gibbons-Hawking term and ${\cal
 L}_{ct}$ is a counter-term that subtracts the infinite terms in the on-shell action and the Brown-York tensor. We will discuss the
 counter-term later. The Gibbons-Hawking term is added in such a manner that $\delta S$ does not depend on the normal derivative
 of $\gamma^{\mu\nu}$.

 We begin by studying the higher-derivative formulation. In this
 case,
 \be
 \delta \int_V f(\ri)=\int_V \Xi_{\mu\nu}\delta {\cal G}^{\mu\nu}+\delta S^1_B+\delta S^2_B,
 \label{1st-step}
 \ee
 where \cite{Waldram},
 \be
 \Xi_{\mu\nu}={f}_\mu^\alpha R_{\nu\alpha}-\frac{1}{2}f{\cal G}_{\mu\nu}+\frac{1}{2}\left({\cal G}_{\mu\nu}\nabla_\alpha\nabla_\beta+
 {\cal G}_{\alpha\mu}{\cal G}_{\beta\nu}\Box-{\cal G}_{\alpha\nu}\nabla_\beta\nabla_\mu-{\cal
 G}_{\alpha\mu}\nabla_\beta\nabla_\nu\right){f}^{\alpha\beta},
 \label{EOM-general}
 \ee
 in which, $\Box=\nabla^\mu\nabla_\mu$. Henceforth, we drop the first term on the right hand side of Eq.\eqref{1st-step}.
 The surface terms are:
 \bea
 \label{deltaS1-1}
 \delta S^1_B&=&-\frac{1}{2}\int_B {f}^{\nu\alpha}\left[n^\beta(\nabla_\nu\delta {\cal G}_{\alpha\beta}+\nabla_\alpha\delta
 {\cal G}_{\beta\nu}-\nabla_\beta\delta {\cal G}_{\nu\alpha})-n_\alpha {\cal G}^{\beta\sigma}\nabla_\nu\delta
 g_{\beta\sigma}\right]\\
 \label{deltaS1-2}
 &=&\int_B \left(f^{\mu\nu}\delta K_{\mu\nu}+s\,\delta
 K\right)+\frac{1}{2}\int_B h^\mu \gamma^{\rho\sigma}{\cal D}_\mu\delta {\cal G}_{\rho\sigma},
 \eea
 where, inspired by Eq.\eqref{9-11-12-n1}, we have defined,
 \be
 s=n_\mu n_\nu f^{\mu\nu},\quad h^\mu=\gamma^\mu_{\ \nu}\,H^\nu,\quad H^\mu=n_\nu f^{\mu\nu},
 \ee
 and,
 \be
 \delta S^2_B=\frac{1}{2}\int_B\left[(n_\nu\delta {\cal G}_{\sigma\alpha}+n_\alpha\delta {\cal G}_{\sigma\nu}-
 n_\sigma\delta {\cal G}_{\nu\alpha})\nabla^\sigma-g^{\sigma\beta}n_\nu\delta {\cal G}_{\sigma\beta}\nabla_\alpha\right]{f}^{\nu\alpha}.
 \ee
 Thus, on-shell,
 \be
 \delta \int_V f(\ri)=\delta {\tilde S}^1_B+\delta {\tilde S}^2_B,
 \ee
 in which,
 \be
 \delta {\tilde S}^1_B=\int_B \left(f^{\mu\nu}\delta K_{\mu\nu}+s\,\delta K\right),
 \ee
 and,
 \be
 \delta {\tilde S}^2_B=\delta {S}^2_B-\frac{1}{2}\int_B \gamma^{\mu\nu}\delta \gamma_{\mu\nu}{\cal D}_\kappa h^\kappa.
 \ee
 The generalized Gibbons-Hawking term should be added such that it remove
 the $n^\alpha\partial_\alpha\delta\gamma_{\mu\nu}$-dependent terms in  $\delta {\tilde
 S}^1$. A  covariant choice is,
 \be
 S_{GH}=-\int_B \left(f^{\mu\nu} K_{\mu\nu}+s K\right).
 \label{Gibbons-Hawking-Hohm-Tonni}
 \ee
 This term has been derived in \cite{Hohm-Tonni} for $D=3$ massive gravity.

 Using  the  normal coordinates,
 \be
 ds^2=N^2(r)dr^2+\gamma_{ij}(r,x^k)\,dx^i dx^j,
 \ee
 the Brown-York tensor is defined by,
 \bea
 T^{ij}&=&\left.\frac{2}{\sqrt{\gamma}}\frac{\delta S}{\delta
 \gamma_{ij}}\right|_{\rm on-shell}=T_1^{ij}+T_2^{ij}+T^{ij}_{\rm{ct}}.
 \eea
  $T^{ij}_{\rm{ct}}$ comes from the counter-terms, to be
discussed later,
 \be
 T_1^{ij}=
 -\frac{2}{\sqrt{\gamma}}\frac{1}{\delta \gamma_{ij}}
 \left(\int_BK_{ab}\delta (f^{ab}\sqrt{\gamma})+K\delta(s\sqrt{\gamma})\right),
 \ee
 and,
 \bea
 \label{1st-term-in-tildeS}
 T_2^{ij}&=&\frac{2}{\sqrt{\gamma}}\frac{\delta {\tilde S}^2_B }{\delta \gamma_{ij}}\nn\\&=&n_\nu \nabla^{(i} f^{j)\nu }
 -n.\nabla f^{ij}-\gamma^{ij}(n_\nu\nabla_\alpha f^{\alpha\nu}+{\cal
 D}_k h^{k})\\
  \label{2nd-term-in-tildeS}
 &=&-K^{(j}_k f^{i)k}-n.\nabla f^{ij}+\gamma^{ij}K_{ab}
 f^{ab}+\nabla^{(i} H^{j)}-\gamma^{ij}(\nabla_\alpha H^\alpha+{\cal D}_k h^k),
 \eea
 in which,
 \bea
 \nabla^i{ H}^j&=&{\cal D}^i {h}^j+K^{ij}s,\nn\\
 \nabla_\alpha { H}^\alpha&=&{\cal D}_a h^a+({\cal D}_r+K)s.
 \eea
 where ${\cal D}_r=n^\mu\partial_\mu$. Furthermore,
 $ \gamma_{ij,r}=2n_rK_{ij}$ and consequently,
 \be
 n.\nabla f^{ij}={\cal D}_r
 f^{ij}+K^{(i}_kf^{j)k}.
 \ee

 So far, our results  are valid in both the second-order
 and the higher-derivative formulations.  If one assumes that $\left.\delta f^\mu_\nu\right|_B=0$ which is a legitimate assumption in the second-order formulation, then,
 \be
 \delta f^{ij}=f^i_{\ k}\,\delta\gamma^{kj},\hspace{1cm}\delta s=0.
 \ee
 In this case,
 \be
 T_1^{ij}=K_k^{(i}f^{j)k}-\left(K_{ab}f^{ab}+sK\right)\gamma^{ij},
 \ee
 and $T_1+T_2$ reproduces the stress-tensor  derived in \cite{Hohm-Tonni} for the massive
 gravity.

 On the contrary, the assumption $\left.\delta f^\mu_\nu\right|_B=0$ can not
 be taken for granted in the higher-derivative formulation of the $f(\ri)$ model, and the contribution from $\left.\delta R^\mu_\nu\right|_B$ has to be  taken into account \cite{Cremonini,Nojiri}.

 Since we are interested in asymptotically locally AdS
 spacetimes, we simplify the problem by assuming  that,
  \be
 \left.{f}^{\mu\nu}\right|_B={\Omega}\, {\cal G}^{\mu\nu},
 \label{assumption}
 \ee
 where, $\Omega$ is a constant.
  In this case, in the both formulations, $\delta {\tilde S}^2$ does not contribute in the Brown-York tensor,
  i.e. $T_2^{ij}=0$,  as can be verified by evaluating Eq.\eqref{1st-term-in-tildeS}.
  Furthermore, the Gibbons-Hawking term \eqref{Gibbons-Hawking-Hohm-Tonni} simplifies to
 \be
 S_{GH}=-2\int_B \Omega (K+\ell^{-1}),
 \label{the-new-new-GH-term-1}
 \ee
 where, we have added  a counter-term similar to Eq.\eqref{GH-Eintein-counter-term}.  Noting that for
 such backgrounds,
 \be
 \left.\frac{d\, {f}^{\mu}_\alpha}{d
 R^\beta_\nu}\right|_B=\Upsilon_1\delta^{\mu}_\alpha\delta^\nu_\beta+\Upsilon_2\delta^{\nu}_\alpha\delta^\mu_\beta,
 \label{upsilon-1}
 \ee
 which follows from Eq.\eqref{f-zegond}, one verifies that,
 \be
  \delta\Omega=\frac{1}{d+1}\delta^\nu_\mu\,\delta{f}^\mu_\nu  =\Upsilon\delta R,\quad  \Upsilon=\Upsilon_1+\frac{\Upsilon_2}{d+1}.
 \label{upsilon-2}
 \ee
 Thus,
 \be
 \delta S=-\int_B\Omega\,t_{ij}\,\delta\gamma^{ij}-2\int_B
 \Upsilon (K+\ell^{-1})\,\delta R,
 \label{0-beg-00}
 \ee
  in which, $t_{ij}$ is defined in Eq.\eqref{Sk-t-ij}.  $\delta R$ is given by Eq.\eqref{deltaR-general} and depends on
  the normal derivative of $\delta \gamma_{ij}$. In principle, one seeks a surface term which removes this term.  In \cite{Madsen-Barrow} it is argued that no such surface term exists in general. In the next section we obtain the corresponding surface term for the asymptotically locally AdS$_3$ spacetimes given by Eq.\eqref{int-1}.
  %%%%%%%%%%%%%%%%%%%%%%%%%%%%%%%%%%%%%%%%%%%%%%%%%%%%%%%%%%%%%%%%%%%%%%%%%%%%%%%%%%%%
 \subsection{Asymptotically locally AdS Einstein solutions}\label{sec-ALAdS}
 Henceforth,  we restrict ourselves to backgrounds which asymptote to locally AdS$_{3}$
 solution, and use the traditional Fefferman-Graham
 asymptotic expansion of the metric given by Eqs.\eqref{Sk-asym-metric} and \eqref{Sk-Fefferman}.\footnote{For an asymptotically AdS spacetime, the metric asymptotes to the exact AdS metric at the boundary. In an asymptotically locally AdS spacetime, the boundary metric is treated as a free field and one can use  the Fefferman-Graham expansion. It is known that  in general, this expansion is insufficient to describe all solutions of $f(\ri)$ models, in particular at the critical point $ \Omega=0$; see \cite{Cunliff} and references therein.}

 In the Fefferman-Graham coordinates, $\delta R$  is given by Eq.\eqref{deltaR2}. In this case, the unwanted term
 in Eq.\eqref{0-beg-00} is encapsulated in $\delta {\cal P}$. Furthermore,
 \be
 K=-\frac{2}{\ell}+{\cal O}(r),
 \ee
 i.e.  $K$ is constant  on the boundary located at $r=0$. Consequently,
 one can use the following counter-term in order to remove  $\delta {\cal
 P}$ in Eq.\eqref{0-beg-00},
  \be
 S^b_\alpha=-\frac{2}{\ell}\int_B\Upsilon{\cal P_\alpha},\hspace{1cm}{\cal P}_\alpha=(1-\alpha)\,{\cal R}+{\cal P}.
 \label{Sub-scheme}
 \ee
 where $\alpha\in{\mathbb R}$ is arbitrary, and $\Upsilon$ is defined in Eqs.\eqref{upsilon-1} and \eqref{upsilon-2}. Note that  ${\cal P}_\alpha\sim{\cal O}(r)$ and ${\cal P}_{\alpha,\,\rm on-shell}=-\alpha\,r\,{\cal R}^{(0)}+{\cal O}(r^2)$. This changes Eq.\eqref{0-beg-00} to,
 \be
  \delta S=-\int_B\Omega\,t_{ij}\,\delta\gamma^{ij}+\frac{2}{\ell}\int_B \Upsilon
  \left(\alpha\,\delta{\cal R}+\frac{1}{2}{\cal P}_{\alpha}\,\gamma_{ij}\,\delta\gamma^{ij}\right)-
  \frac{2}{\ell}\int_B{\cal P}_{\alpha}\,\delta\Upsilon.
  \label{0-beg}
  \ee
   Eqs.\eqref{particular} and \eqref{deltaR2} imply that ${\cal P}_\alpha\,\delta\Upsilon\sim{\cal O}(r^2)$  and consequently, the last term in Eq.\eqref{0-beg} is vanishing.\footnote{Recall that $\delta\gamma^{ij}\sim{\cal O}(r)$ and
   $\sqrt{\gamma}\sim{\cal O}(r^{-1})$.} Therefore, no further counter-term is needed in order to make the variational principle well-defined. The second term in Eq.\eqref{0-beg} is  vanishing on-shell because,
   \bea
 \int_B \gamma^{ij}\delta{\cal
 R}_{ij}=\int_B\left({\cal D}_i{\cal D}_j -\gamma_{ij}{\cal D}^k{\cal D}_k\right)\,\delta
 \gamma^{ij}=0,
 \eea
  and,
 \be
 {\cal R}_{ij}=\frac{1}{2}{\cal R}^{(0)} g^{(0)}_{ij}+{\cal O}(r).
  \ee
  In summary, we have verified that the variational principle is well-defined for the action,
  \be
 S=\int_V f-2\int_B \Omega (K+\ell^{-1})-\frac{2}{\ell}\int_B\Upsilon{\cal P}_{\alpha},
  \label{final-action}
  \ee
  and the corresponding Brown-York tensor is,
  \be
  {T}_{ij}^{\rm ren}=2\,\Omega\,t_{ij}.
  \label{stress-tensor-1}
  \ee

  We still need to  determine another counter-term  which subtracts the logarithmic divergence in the on-shell value of the
 action \eqref{final-action} \cite{deHaro}. Using Eq.\eqref{squart-G} one obtains,
 \be
 \left.\int_V f\right|_{\rm on-shell}=\lim_{\epsilon\to0}\frac{\ell
 f_0}{2}\int d^2x\sqrt{g^{(0)}}\left(\frac{1}{\epsilon}-\frac{{\rm tr} g^{(2)}}{2}\ln
 \epsilon\right)+\mbox{finite},
 \label{epsilon-1-f}
 \ee
 where, $f_0$ denotes the (asymptotic) on-shell value of $f(\ri)$. Furthermore,
 \be
 -2\left.\int_B\Omega\,(K+\ell^{-1})\right|_{\rm on-shell}=\frac{2\,\Omega}{\ell}\int d^2x
 \sqrt{g^{(0)}}\epsilon^{-1}+{\rm finite},
 \label{epsilon-1-K}
 \ee
 and,
 \be
 -\frac{2}{\ell}\left.\int_B\Upsilon\,{\cal P}_{\alpha}\right|_{\rm
 on-shell}=\frac{2}{\ell}\left(4\pi\chi\right)\Upsilon\,\alpha=\mbox{finite},
 \label{difference}
 \ee
 where, $\chi$ is the Euler characteristic of the boundary  given by Eq.\eqref{Euler}.
 For an AdS$_3$ solution, $\Omega$ in Eq.\eqref{assumption} is a constant, and the equation of motion $\Xi_{\mu\nu}=0$ implies that
 \be
 f_0+4\ell^{-2}\Omega=0.
 \label{f0-omega0}
 \ee
 Consequently, the $\epsilon^{-1}$-terms in Eqs.\eqref{epsilon-1-f} and \eqref{epsilon-1-K} cancel out, and,
 \be
 S_{\rm on-shell}=-\frac{\ell\,\Omega}{2}(4\pi\chi)\lim_{\epsilon\to0}\ln
 \epsilon+\mbox{finite}.
 \label{log-div}
 \ee

 The parameter $\alpha$ in Eq.\eqref{Sub-scheme} remains arbitrary. This reflects the fact that classically, one can arbitrarily add or remove the Euler characteristic to the action. Since this is a finite term, holographic renormalization is also ignorant of it. In principle, $\alpha$ can be determined by AdS/CFT correspondence, since  the on-shell value of the action gives the leading term in the CFT partition function \cite{witten}.

 The formula \eqref{0-beg} is obtained in the higher-derivative formulation given by the action \eqref{higher-d-action}.
 By simply omitting the $\Upsilon$-terms, one obtains the corresponding formula in the second-order formulation \eqref{second-order-action}.
  \subsection*{Brown-York stress-tensor}
  Since the log-counter-term is a topological term, it will not contribute to the Brown-York stress tensor \eqref{stress-tensor-1}.
 Using Eq.\eqref{Sk-div-of t} one verifies that,
 \be
 {\cal D}^i T_{ij}^{\rm ren}=0.
 \label{b}
 \ee
  Furthermore,
 \be
  {\rm tr}T^{\rm ren}=\ell\,{\Omega}\,{\cal R}^{(0)}=\frac{c}{24\pi}{\cal R}^{(0)}.
 \label{new-trace-anomaly1}
 \ee
  Thus,
 \be
 c=\frac{3\ell}{2G}(16\pi G\, \Omega),
 \ee
 which is the central charge obtained  in \cite{Hohm-Tonni,Kraus-Larsen}.  Eq.\eqref{log-div} implies that, similar to Eq.\eqref{Sk-chi}, ${\rm tr}T$ is given by the logarithmic divergence of the action \cite{Henningson-Skenderis,deHaro,Kraus-Larsen},
 \be
 S_{\log-{\rm term}}=-\frac{c\,\chi}{12}\ln\epsilon.
 \label{f-of-r-chi}
 \ee

 %=============================================================================================
 \section{A non-covariant cut-off independent counter-term}\label{Sec-GA}
 By the AdS/CFT correspondence, the leading term in the CFT partition function is given by the finite term of the classical gravity action \cite{witten}
 \be
 \left<\exp\int_B\phi^{(0)}{\cal O}\right>_{\rm CFT}=\exp\left(-S(\phi_{\rm cl})\right),
 \ee
 in which $\phi^{(0)}$ denotes the boundary value of the classical field $\phi_{\rm cl}$,
 and the expectation value of the stress-energy tensor of the dual   CFT is identified with the Brown-York tensor \cite{Bala}.

 The finite term in the gravity action \eqref{log-div} is the sum of the finite terms in the bulk term \eqref{epsilon-1-f} and the boundary terms \eqref{epsilon-1-K} and \eqref{difference}.  The contribution from the  boundary terms  is given by,
 \be
 4\,\pi\,\chi\left(\frac{\ell\,\Omega}{2}+\frac{2\,\alpha\,\Upsilon}{\ell}\right),
 \label{finite-boundary-terms}
 \ee
 where, $\chi$ is the Euler characteristic of the boundary. It is a topological term  and consequently, the boundary data $g^{(0)}_{ij}$ is obscured in this term.

 A closely related problem is the value of the divergence of  the stress-tensor.
 The argument in \cite{Imbimbo} reviewed in section \ref{Sec-PBH}, as well as the method of \cite{Kraus-Larsen} can not determine the divergence of the stress-tensor. Since $f(\ri)$ gravity is parity-preserving, there is no room for a   gravitational anomaly in the dual CFT given
 by,
 \be
 {\cal D}^iT_{ij}=\beta\,\epsilon^i_j\partial_i{\cal R}^{(0)},
 \ee
 i.e.  $\beta=0$.  Nevertheless, one  can still add boundary local terms which induce a gravitational anomaly given by,
 \be
 {\cal D}^iT_{ij}=\frac{b}{24\pi}\,\partial_j{\cal R}^{(0)}.
 \ee
 The holographic renormalization can produce such an anomaly,
 depending on the counter-term one uses to subtract the logarithmic divergence in the on-shell value of the action given by Eqs.\eqref{Sk-chi} and \eqref{f-of-r-chi}.

 The prescription in \cite{Henningson-Skenderis, deHaro} is subtracting the `covariant' cut-off dependent counter-term $S^{\rm ct}_{\rm log}=-S_{\rm log-term}$ given in Eq.\eqref{f-of-r-chi}.  This results in Eq.\eqref{finite-boundary-terms}. One can instead use another counter-term which is independent of the cut-off,
 \be
 S^{\rm ct}_{\rm log}=-\frac{c}{48\pi}\int_B{\cal R}\sqrt{\gamma}\ln\sqrt{\gamma}=
 -\frac{c}{48\pi}\int_B{\cal R}^{(0)}\sqrt{g^{(0)}}\left(-\ln \epsilon+\ln\sqrt{g^{(0)}}\right).
 \ee
 This counter-term is not covariant. Its contribution to the on-shell value of the classical action is
  \be
  -\frac{c}{48\pi}\int_B{\cal R}^{(0)}\sqrt{g^{(0)}}\,\ln\sqrt{g^{(0)}}.
 \ee
  which, unlike the topological term \eqref{finite-boundary-terms} inherits the boundary data. Furthermore, it
adds a new term to the Brown-York tensor,
 \be
 T^{\rm ct}_{{\rm log},\,ij}=\left.-\frac{c}{48\pi}{\cal R}\gamma_{ij}\right|_{r=0}=-\frac{c}{48\pi}{\cal R}^{(0)}g^{(0)}_{ij}.
 \ee
 In this scenario, the renormalized  Brown-York tensor is,
 \be
 T_{ij}=\frac{c}{12\,\pi\ell}\,t_{ij}-\frac{c}{48\pi}{\cal R}^{(0)}g^{(0)}_{ij}.
 \ee
 Consequently,
 \bea
 {\rm tr}\,T&=&0,\nn\\
 {\cal D}^iT_{ij}&=&-\frac{c}{48\pi}\partial_j{\cal R}^{(0)},
 \eea
 which is similar to the case studied in \cite{David}. This observation motivates us to consider a more general
 situation, where,
 \be
 T_{ij}=\frac{(a-2b)}{12\pi\ell}\,t_{ij}+\frac{b}{24\pi}{\cal
 R}^{(0)}g^{(0)}_{ij},
 \label{generalized-BY}
 \ee
 which  gives,
  \be
  \label{anomaly-ab}
  T^i_i=\frac{a}{24\pi}\,\mathcal{R}^{(0)},\hspace{1cm} \nabla_j T^j_i=\frac{b}{24\pi}\,\partial_i \mathcal{R}^{(0)}.
  \ee
 For the covariant subtraction $(a,b)=(c,0)$, and for the cut-off independent subtraction $(a,b)=(0,-c/2)$.
 \subsection{Hawking effect of a 2d Schwarzschild black hole}
 In the following, we show that the true value of the central charge $c=a-2b$ can be recognized via the Hawking effect of an asymptotically flat  two-dimensional black hole located on the boundary \cite{Sol-BH}.   Consider a Schwarzschild black
 hole,
  \be
  ds^2=-u(x)\,dt^2+\frac{dx^2}{u(x)},
  \label{2d-BH}
  \ee
  where $u(x)$ has a simple zero at $x_h$ indicating the event-horizon
  and,
  \be
  \lim_{x\to\infty}u(x)=1.
  \ee
  The non-vanishing Christoffel symbols are,
  \be
  \Gamma^t_{tx}=-\Gamma^x_{xx}=\frac{u'}{2u},\hspace{1cm}\Gamma^x_{tt}=\frac{u u'}{2}.
  \ee
  and $\mathcal{R}^{(0)}=-u''(x)$.  Eq.\eqref{anomaly-ab} reads,
 \be
 \begin{array}{l}
 T^x_x+T^t_t=-\frac{a}{24\pi}u'',\\
 \partial_x T^x_x+\frac{u'}{2u}(T^x_x-T^t_t)=-\frac{b}{24\pi}u''',\\
 \partial_x T^x_t=0.
 \end{array}
 \ee
 These equations can be solved and the integration constants can be determined by requiring that:
 ({\em a}) $T^t_t$ and $T_{xt}$ are finite at the horizon \cite{Sol-BH},
 and ({\em b}) asymptotically,
 \be
 T_{tt}=c_+\frac{\pi}{6}T_H^2,\hspace{1cm} T_{xt}=c_-\frac{\pi}{6}T_H^2,
 \ee
 in which, $T_H=g'(x_+)/4\pi$ is the Hawking temperature of the black hole and $c_\pm=(c_L\pm c_R)/2$. Finiteness of $T_{xt}$ at the horizon implies that $c_-=0$ and consequently no gravitational anomaly is detected by the Hawking effect, i.e. $c_L=c_R$. Finiteness of $T^t_t$ at the horizon gives,
 \be
 c_+=a-2b.
 \label{c-plus}
 \ee
 \subsection{BTZ-black hole}\label{Sec-BTZ}
 It is interesting to note that the true value of the central charge can  also be recognized by studying BTZ black holes.
 The boundary of a static BTZ black hole is a flat torus, i.e.  both the trace-anomaly and the gravitational anomalies
 \eqref{anomaly-ab}  are vanishing in this case. Thus, the BTZ-black hole can be used to verify, via holography, whether $c_+$ defined by Eq.\eqref{c-plus} is the correct central charge or not.

 The BTZ geometry,
 \be
 ds^2=-(r^2-8GM\ell^2)\,dt^2+\frac{\ell^2dr^2}{r^2-8GM\ell^2}+r^2d\phi^2,
 \label{BTZ-metric}
 \ee
  in the Fefferman-Graham coordinates is given by,
 \bea
 g^{(0)}_{ij}&=&\eta_{ij},\nn\\
 g^{(2)}_{ij}&=&4G M\ell^2\delta_{ij}=\frac{2\pi^2\ell^2}{\beta^2}\delta_{ij},
 \eea
 in which, $\eta_{ij}={\rm diag}(-1,1)$, $\delta_{ij}={\rm diag}(1,1)$ and the Hawking temperature $\beta^{-1}$ gives  the torus complex structure  $\tau=i\beta/2\pi$. Since ${\cal R}^{(0)}=-2\ell^{-2}{\rm
 tr}g^{(2)}=0$,  Eq.\eqref{generalized-BY}  gives,
 \be
 T_{ij}=\frac{\pi c}{6\beta^2}\delta_{ij}=-\frac{c}{24\pi}\frac{1}{\tau^2}\delta_{ij}.
 \ee
  To see why this result  is important recall that the CFT free-energy of a BTZ black hole
  can be obtained by a modular transformation $\tau\to-\tau^{-1}$ from the the free-energy of the vacuum which,
  corresponds to the thermal AdS
  \cite{Maldacena:1998bw,Kraus-Larsen},
  \be
 I_{\rm BTZ}(\tau, \bar\tau) = - \frac{i \pi}{12}\left( \frac{c_L}{ \tau} -\frac{c_R}{\bar\tau}\right).
 \ee
 Consequently, the corresponding CFT weights are,
 \bea
 \Delta&=&-\frac{1}{2\pi i}\frac{\partial I}{\partial
 \tau}=-\frac{c_L}{24\tau^2},\nn\\
 \bar\Delta&=&\frac{1}{2\pi i}\frac{\partial I}{\partial
 \bar\tau}=-\frac{c_R}{24{\bar\tau}^2}.
 \eea
 Thus, $\Delta+\bar\Delta$ is equivalent to the Brown-York mass of the black hole,
 \be
 M_{\rm BY}=\int_0^{2\pi} d\phi \,T_{00}=\frac{\pi^2c}{3\beta^2}.
 \ee
 Note that the time coordinate $t$ in Eq.\eqref{BTZ-metric} equals, $\ell^{-1}t_{\rm BTZ}$, and consequently, $M_{\rm BY}=\ell M_{\rm BTZ}$.
 The Cardy formula gives \cite{Strominger-97},
 \be
 S_{\rm Cardy}=2\pi\sqrt{\frac{c_L\Delta}{6}}+2\pi\sqrt{\frac{c_R\bar\Delta}{6}}=\frac{c}{6\ell}{\cal A}_{\rm BTZ},
 \ee
 where ${\cal A}_{\rm BTZ}$ is the area of the event horizon,
 \be
 {\cal A}_{\rm BTZ}=2\pi\ell\left(\frac{2\pi}{\beta}\right).
 \ee

  %=============================================================================================================================

 %===================================== =========================================

 \section{Discussion}\label{Sec-discussion}
  For backgrounds in which,   the traditional Fefferman-Graham expansion is
  available,   we found the Gibbons-Hawking term in the higher-derivative formulation of $f(\ri)$ gravity,
  and determined   the corresponding   counter-terms.
  The resulting Brown-York tensor appeared to be equivalent to the one obtained in the second-order formulation,
  in which,  an auxiliary field is used.

  We also verified that the logarithmic divergence of the on-shell action can be subtracted either by a cut-off
  dependent covariant counter-term quite similar to the one used in \cite{Henningson-Skenderis, deHaro}, or by a
  cut-off independent non-covariant counter-term. In the former case, one obtains a trace anomaly equivalent to the
  one obtained in \cite{Kraus-Larsen,Saida-Soda}. In the later case, the Weyl anomaly is vanishing and  one
  encounters a gravitational anomaly instead, which can be exchanged for the familiar Weyl anomaly by adding a
  local surface term. We verified that, keeping the gravitational anomaly, one can determine
  the value of the central charge in term of the Hawking effect of a Schwarzschild black hole placed on the boundary, or by means of
  BTZ holography.

 The CFT dual to $f(\ri)$ gravity should address various phenomena which are absent in General Relativity.
 For example, the Ostrogradski's theorem implies that  $f(\ri)$ theories are in general instable
 \cite{Woodard}.  From this point of view, $f(R)$ models in which $f$ is an algebraic function of undifferentiated
 Ricci scalar are viable models \cite{Woodard}. Of course, in these models positivity of the
 screened Newton's constant requires  that  $\Omega\sim f'>0$.
 This condition is also necessary for the unitarity of the boundary CFT as it implies that the central charge given by the holographic Weyl anomaly
  is positive.  Unitary $f(\ri)$ gravities in three dimensions and their CFT duals are widely  studied,
 see e.g. \cite{Tekin} and references therein.

 In the context of $f(R)$ gravity, Ricci stability also imposes $\Upsilon\sim f''(R)>0$
 \cite{Sotiriou}, which should be addressed in the dual CFT.  Furthermore, there is vDVZ discontinuity
 \cite{vDVZ} in $f(R)$ gravity models \cite{Myung} since $f(R)$ gravity models are essentially equivalent
 to GR with an additional scalar.  Thus, it is necessary to realize the vDVZ discontinuity in the CFT dual. We could not trace these effects in the holographic renormalization of the theory, since both the Brown-York stress-tenor and the on-shell  action appeared to be insensitive to such details.
  %===================================== =========================================

 \appendix

 \section{$f(\ri)$ as a Polynomial in $\ri$}
 In this appendix, we compute $f^{\mu}_\nu$ and $d\,f^\alpha_\beta/d\ri$. Assuming that $f$ is a polynomial in $\ri$,
 \be
 f(\ri)=\sum_{\{n_1\cdots n_k\}}c_{n_1\cdots n_k}R^{n_1}\cdots R^{n_k},
 \ee
 where,
 \be
 R^n=\left({R^n}\right)^\mu_{\  \mu},\quad \left({R^{n+1}}\right)^\mu_{\ \nu}=R^\mu_{\ \alpha_1}R^{\alpha_1}_{\ \alpha_2}\cdots R^{\alpha_n}_{\
 \nu},\quad  1^\mu_{\ \nu}=\delta^\mu_\nu,
 \ee
 one verifies that,
 \be
 \delta f=\sum_{\{n_1\cdots n_k\}}c_{n_1\cdots n_k}\sum_{i=1}^k R^{n_1}\cdots \delta R^{n_i}\cdots R^{n_k},
 \ee
 in which,
 \be
 \delta R^{n}=n\left(R^{n-1}\right)^\mu_{\ \nu}\delta R^\nu_{\ \mu}.
 \label{delta-R-n-1}
 \ee
 Thus,
 \be
 {f}^\beta_{\ \alpha}=\sum_{\{n_1\cdots n_k\}}c_{n_1\cdots n_k}\sum_{i=1}^k n_i\,R^{n_1}\cdots \widehat{R^{n_i}}\cdots R^{n_k}\left(R^{n_i-1}\right)^\beta_{\ \alpha},
 \ee
  where, the term with a hat is replaced by 1, e.g. $ x\widehat{y}z=xz$.
  In order to compute $  \delta {f}^\beta_{\ \alpha}$  one needs to
  compute,
  \be
 \delta \left[\left(\prod_{i=1}^k R^{n_i}\right)\left(R^{m}\right)^\beta_{\ \alpha}\right],
  \ee
  which is given by Eq.\eqref{delta-R-n-1} and,
  \be
  \delta\left(R^n\right)^\nu_{\ \mu}=\sum_{k=0}^{n-1} \left(R^k\right)^\alpha_{\ \mu}\left(R^{n-k-1}\right)^\nu_{\ \beta}\delta R^{\beta}_{\ \alpha}.
  \ee
  Consequently,
  \bea
  \frac{d\, {f}^\beta_{\ \alpha}}{d R^\rho_{\ \sigma}}&=&\sum_{\{n_1\cdots n_k\}}c_{n_1\cdots n_k}\Big[\nn\\
  &&\sum_{i\neq j} n_in_j\,R^{n_1}\cdots \widehat{R^{n_j}}\cdots\widehat{R^{n_i}}\cdots R^{n_k}\left(R^{n_j-1}\right)^\sigma_{\ \rho}\left(R^{n_i-1}\right)^\beta_{\ \alpha}\nn\\
  &+&\sum_i n_i R^{n_1}\cdots \widehat{R^{n_i}}\cdots R^{n_k}\sum_{j=0}^{n_i-2}\left(R^j\right)^{\sigma}_{\ \alpha}\left(R^{n_i-j-2}\right)^{\beta}_{\ \rho}\Big].
  \label{f-zegond}
  \eea

 \section{Induced geometry on the boundary}
 In this paper, we assume that the spacetime given by the metric
 ${\cal G}_{\mu\nu}$ is surrounded by a space-like  boundary $B$ given by a continuous and surface-forming vector
 field $n^\mu$ \cite{Madsen-Barrow},
 \be
 n_\mu n^\mu=1,\hspace{1cm}\nabla_{[\alpha}n_{\beta]}=0.
 \label{n-def}
 \ee
 Furthermore, we assume that this vector field is `inward' pointing  normal to the
 boundary.  The induced metric on the boundary is given by,
 \be
 \gamma_{\mu\nu}={\cal G}_{\mu\nu}-n_\mu n_\nu,
 \label{9-11-12-n1}
 \ee
 where,
 \be
 n^\mu\gamma_{\mu\nu}=0,\quad\gamma_{\mu\rho}\gamma^{\rho\nu}=\delta_\mu^\nu-n_\mu n^\nu.
 \ee
 The extrinsic curvature of the boundary is defined by
 \be
 K_{\mu\nu}=\nabla_\mu n_\nu.
 \label{10-11-12-m4}
 \ee
   It is useful to recall that  in the ADM decomposition,
 \be
 ds^2=N^2dr^2+\gamma_{ij}(dx^i+N^idr)(dx^j+N^jdr),
 \ee
 $n_\rho=(0^i,N)$ and $n^\rho=(-N^{-1}N^i,N^{-1})$. Furthermore,
 $\delta {\cal G}_{ij}=\delta \gamma_{ij}$ and $\delta {\cal G}_{ri}=N^j\delta\gamma_{ij}$.
 One defines the Brown-York tensor with respect to $\delta\gamma_{ij}$, assuming that $\delta N=0$. It is clear
 that $\delta n_\rho=0$ and $n^\rho\delta {\cal G}_{\rho i}=0$. See also
 appendix A of \cite{Hohm-Tonni}.

 Following section 10 of \cite{Waldbook} and noting that here, $\gamma_{\mu\nu}=g_{\mu\nu}-n_\mu n_\nu$ i.e. $n^2=1$ ,
 one verifies that,
  \be
 {\cal R}_{ijk}^{\ \ \ \ l}=\gamma_i^\mu \gamma_j^\nu \gamma_k^\rho \gamma^l_\sigma R_{\mu\nu\rho}^{\ \ \ \
 \sigma}+K_{ik}K_j^l-K_{jk}K_i^l,
 \label{Walds-Identity}
 \ee
 where, ${\cal R}_{ijk}^{\ \ \ \ l}$ denotes the Riemann tensor defined with respect to $\gamma_{ij}$, the metric induced on the  boundary. Similar to \cite{Waldbook}, our curvature convention is
   $[\nabla_\rho,\nabla_\sigma]A^\mu=R^\mu_{\ \nu\rho\sigma}A^\nu$ and $R_{\mu\nu}=R^\rho_{\ \mu\rho\nu}$.
 Consequently,
 \be
  {\cal R}_{ij}=\gamma^{\mu}_i\gamma^\nu_j(R_{\mu\nu}-n^\alpha n^\beta R_{\mu\alpha\nu
  \beta})+KK_{ij}-K_{ik}K_j^k.
 \ee
  Since, using Eq.\eqref{n-def},
 \be
 n^\alpha n^\beta R_{\mu\alpha\nu \beta}=n^\beta [\nabla_\nu,\nabla_\beta]n_\mu=-K^\beta_\nu K_{\beta\mu}-n.\nabla
 K_{\mu\nu},
 \label{yadam1}
 \ee
 one verifies that,
 \be
 {\cal R}_{ij}=\gamma^{\mu}_i\gamma^\nu_j R_{\mu\nu}+n.\nabla
 K_{ij}+KK_{ij}.
 \label{Ricci-two-dim-off-shell}
 \ee
 This gives, in particular \cite{Madsen-Barrow},
  \be
  {\cal R}=R+K_{ij}K^{ij}+K^2+2n.\nabla K,
   \label{Key-R-identity}
   \ee
  where, we have used,
  \be
  R_{\mu\nu}n^\mu n^\nu=n^\nu[\nabla_\rho,\nabla_\nu]n^\rho=-K_{\mu\nu}K^{\mu\nu}-n^\mu\partial_\mu
  K.
   \ee

 In order to obtain the Gibbons-Hawking term, one needs to compute
 the surface terms that appear in the  variation of the action with respect to the  metric. Assuming that  the
 boundary  $B$ is fixed \cite{Madsen-Barrow}, i.e. $\delta n_\rho=0$ and $\delta \gamma_{\mu\nu}$ is tangential,
 \be
 n^\mu\delta \gamma_{\mu\nu}=0,\quad\delta n^\rho=0,\quad \gamma_{\mu\nu}\delta
 \gamma^{\mu\nu}=-\gamma^{\mu\nu}\delta \gamma_{\mu\nu}.
 \label{boundary-variation}
 \ee
 one obtains, using Eq\eqref{10-11-12-m2},
  \be
   \label{deltaKij}
    \delta K_{\mu\nu}=-\frac{1}{2}n^\rho\left(\nabla_\mu\delta
  \gamma_{\nu\rho}+\nabla_\nu\delta \gamma_{\mu\rho}-\nabla_\rho\delta
  \gamma_{\mu\nu}\right).
  \ee
  Consequently,
  \bea
  \label{deltaK}
    \delta K&=&\frac{1}{2}\,\gamma^{\mu\nu}n^\rho\nabla_\rho\delta
  \gamma_{\mu\nu},\\
  \label{test-2}
  \delta (K^{\mu\nu}K_{\mu\nu})&=&K^{\mu\nu}n^\rho\nabla_\rho \delta
  \gamma_{\mu\nu},\\
  \label{test-3}
  \delta(2\,n^\mu\partial_\mu K)&=&\gamma^{\mu\nu}n^\alpha n^\beta
  \nabla_\alpha \nabla_\beta \delta \gamma_{\mu\nu}.
  \eea
  Therefore,
  \be
  \left.\delta R\right|_B=\delta{\cal R}-\left[(K^{\mu\nu}+K\,\gamma^{\mu\nu})(n.\nabla)+
  \gamma^{\mu\nu}(n.\nabla)^2\right]\delta \gamma_{\mu\nu}.
  \label{deltaR-general}
  \ee

 %===============================================================================================
 \section{Curvature in Fefferman-Graham coordinates}
 The asymptotically locally AdS$_3$ backgrounds,
  \be
 ds^2=\frac{\ell^2}{4}\left(\frac{dr}{r}\right)^2+\gamma_{ij}\,dx^i
 dx^j,\quad\gamma_{ij}=r^{-1}g_{ij},
 \label{Gmetric}
 \ee
 where, the boundary is located  at $r=0$, can be given in terms of the Fefferman-Graham expansion \cite{Henningson-Skenderis,Fefferman-Graham-paper},
 \be
 g_{ij}=\sum_{n=0}^{d/2}g^{(2n)}_{ij}(x)\,r^n+h^{(d)}_{ij}r^{d/2}\ln r+{\cal
 O}(r^{d/2+1}).
 \ee
 For $d=2$ one obtains,
 \bea
 \label{9-11-12-2}
 K_{ij}&=&-\frac{1}{\ell r}\left(g^{(0)}_{ij}-rh_{ij}+{\cal O}(r^2)\right),\\
  \label{9-11-12-3}
 K&=&\ell^{-1}\left\{-2+r\left[{\rm tr}g^{(2)}+(1+\ln r){\rm tr}h\right]+{\cal
 O}(r^2)\right\},
 \eea
 where, the trace operator is defined with respect to $g^{(0)}_{ij}$; e.g. ${\rm tr} h={g^{(0)}}^{ij}h_{ij}$, and we have used the
 equality,
 \be
  \gamma^{ij}=r\left({g^{(0)}}^{ij}-r{g^{(2)}}^{ij}-r\ln r h^{ij}+{\cal
  O}(r^2)\right),
   \label{9-11-12-4}
 \ee
 in which, the $i$ and $j$ indices  are raised and lowered by $g^{(0)}_{ij}$.
  Some other useful identities are:
 \bea
 K^2&=&4\,\ell^{-2}\left\{1- r\left[{\rm tr}g^{(2)}+(1+\ln r)\,{\rm tr} h\right]\right\}+{\cal O}(r^2),\nn\\
 K_{ij}K^{ij}&=&2\,\ell^{-2}\left\{1- r\left[{\rm tr}g^{(2)}+(1+\ln r)\,{\rm tr} h\right]\right\}+{\cal O}(r^2),\nn\\
 n.\nabla K&=& 2\,r\,\ell^{-2}\left[{\rm tr}g^{(2)}+(2+\ln r)\,{\rm tr}h\right]+{\cal
 O}(r^2),
 \eea
 which, using Eq.\eqref{Key-R-identity} give,
 \be
 R=-\frac{6}{\ell^2}+r{\cal R}^{(0)}+\mathcal{P}+{\cal O}(r^2),\hspace{1cm}{\cal
 P}=\frac{6}{\ell}\left(K+\frac{2}{\ell}\right)-2\,n.\nabla K\sim{\cal
 O}(r).
 \label{particular}
 \ee
 Here,  ${\cal R}^{(0)}$ denotes the scalar curvature defined with respect to $g^{(0)}_{ij}$. One
 verifies that,
 \be
 {\cal R}_{ij}={\cal R}^{(0)}_{ij}+{\cal O}(r),
 \label{10-11-12-m5}
 \ee
 where, ${\cal R}^{(0)}_{ij}$ is the Ricci tensor corresponding to
 $g^{(0)}_{ij}$ and consequently,
 \be
 {\cal R}=r{\cal R}^{(0)}+{\cal O}(r^2).
 \ee
 Thus,
 \be
 \delta R=r\,\delta {\cal R}^{(0)}+\delta \mathcal{P}+{\cal O}(r^2).
 \label{deltaR2}
 \ee
 Since ${\rm tr}h=0$ on-shell  \cite{deHaro}, one obtains,
 \be
 {\cal P}_{\rm on-shell}=\frac{2\,r}{\ell^2}{\rm tr} g^{(2)}+{\cal O}(r^2)=-r{\cal R}^{(0)}+{\cal O}(r^2).
 \label{P-onshell}
 \ee

 %==============================================================

 %==============================================================
 \section{Some useful identities}
 In sections \ref{Sec-Einstein} and \ref{sec-Ricci-tensor}, we have used the following identities,
 \be
 \delta \Gamma_{\alpha\beta\rho}=\frac{1}{2}\left(\nabla_\rho\delta {\cal G}_{\alpha\beta}+\nabla_\beta\delta {\cal G}_{\alpha\rho}-
 \nabla_\alpha\delta {\cal G}_{\beta\rho}\right)+\Gamma^{\sigma}_{\rho\beta}\delta {\cal
 G}_{\sigma\alpha},
 \ee
 where, $\Gamma_{\alpha\beta\rho}={\cal G}_{\alpha\sigma}\Gamma^\sigma_{\beta\rho}$.
 Consequently,
 \be
 \delta
 \Gamma^\sigma_{\beta\rho}=\frac{1}{2}{\cal G}^{\sigma\alpha}\left(\nabla_\rho\delta {\cal G}_{\alpha\beta}+\nabla_\beta\delta {\cal G}_{\alpha\rho}-\nabla_\alpha\delta {\cal G}_{\beta\rho}\right).
 \label{10-11-12-m2}
 \ee
 This identity can be used to show that,
 \be
  {\cal G}^{\mu\nu}\delta R_{\mu\nu}={\cal G}^{\mu\nu}\left(\nabla_\rho\delta\Gamma^\rho_{\mu\nu}-\nabla_\nu\delta\Gamma^\rho_{\rho\mu}\right)
 =(-\nabla_\mu\nabla_\nu +{\cal G}_{\mu\nu}\Box)\,\delta {\cal G}^{\mu\nu}.
 \label{10-11-12-m3}
 \ee

 %========================================================================================================================

 %==========================================================================================================================
 

\begin{thebibliography}{99}
 %---------------------------------------------------------------------------------------------------------------
 \bibitem{BTZ1}
  M.~Banados, C.~Teitelboim and J.~Zanelli,
  %``The Black hole in three-dimensional space-time,''
  Phys.\ Rev.\ Lett.\ {\bf 69} 1849 (1992),
  [arXiv:hep-th/9204099].
  %%CITATION = PRLTA,69,1849;%%

\bibitem{BTZ2}
  M.~Banados, M.~Henneaux, C.~Teitelboim and J.~Zanelli,
  %``Geometry of the (2+1) black hole,''
  Phys.\ Rev.\  {\bf D48} 1506 (1993),
  [arXiv:gr-qc/9302012].
  %%CITATION = PHRVA,D48,1506;%%

  \bibitem{Maldacena:1998bw}
  J.~M.~Maldacena and A.~Strominger,
  %``AdS(3) black holes and a stringy exclusion principle,''
  JHEP {\bf 9812}, 005 (1998)
  [hep-th/9804085].

  \bibitem{Cardy}
  J.~L.~Cardy,
 % ``Operator Content of Two-Dimensional Conformally Invariant Theories,''
  Nucl.\ Phys.\  B {\bf 270}, 186 (1986).
  %%CITATION = NUPHA,B270,186;%%

  \bibitem{Strominger-97}
 A.~Strominger,
 % ``Black hole entropy from near horizon microstates,''
  JHEP {\bf 9802}, 009 (1998)
  [arXiv:hep-th/9712251].
  %%CITATION = JHEPA,9802,009;%%

  \bibitem{Brown-Henneaux} J.~D.~Brown and M.~Henneaux,
  %``Central Charges in the Canonical Realization of Asymptotic Symmetries: An Example from Three-Dimensional Gravity,''
  Commun.\ Math.\ Phys.\  {\bf 104}, 207 (1986).
  %%CITATION = CMPHA,104,207;%%

 \bibitem{Henningson-Skenderis}M.~Henningson and K.~Skenderis,
  %``The Holographic Weyl anomaly,''
  JHEP {\bf 9807}, 023 (1998)  [hep-th/9806087]; M.~Henningson and K.~Skenderis,
  %``Holography and the Weyl anomaly,''
  Fortsch.\ Phys.\  {\bf 48}, 125 (2000)  [hep-th/9812032].

 \bibitem{deHaro} S.~de Haro, S.~N.~Solodukhin and K.~Skenderis,
  %``Holographic reconstruction of space-time and renormalization in the AdS / CFT correspondence,''
  Commun.\ Math.\ Phys.\  {\bf 217}, 595 (2001)
  [hep-th/0002230].

  \bibitem{Bala}
  V.~Balasubramanian and P.~Kraus,
  %``A Stress tensor for Anti-de Sitter gravity,''
  Commun.\ Math.\ Phys.\  {\bf 208}, 413 (1999)
  [hep-th/9902121].

  \bibitem{Brown-York} J.~D.~Brown and J.~W.~York, Jr.,
  %``Quasilocal energy and conserved charges derived from the gravitational action,''
  Phys.\ Rev.\ D {\bf 47}, 1407 (1993)  [gr-qc/9209012].

  \bibitem{NMG} E.~A.~Bergshoeff, O.~Hohm and P.~K.~Townsend,
  %``Massive Gravity in Three Dimensions,''
  Phys.\ Rev.\ Lett.\  {\bf 102}, 201301 (2009)
  [arXiv:0901.1766 [hep-th]]; E.~A.~Bergshoeff, O.~Hohm and P.~K.~Townsend,
  %``More on Massive 3D Gravity,''
  Phys.\ Rev.\ D {\bf 79}, 124042 (2009)
  [arXiv:0905.1259 [hep-th]].

  \bibitem{witten}  E.~Witten,
    %``Anti-de Sitter space and holography,''
  Adv.\ Theor.\ Math.\ Phys.\  {\bf 2}, 253 (1998)
  [hep-th/9802150].

  \bibitem{Gibbons-Hawking}G.~W.~Gibbons and S.~W.~Hawking,
  %``Action Integrals and Partition Functions in Quantum Gravity,''
  Phys.\ Rev.\ D {\bf 15}, 2752 (1977).

  \bibitem{Madsen-Barrow} M.~S.~Madsen and J.~D.~Barrow,
  %``De Sitter Ground States And Boundary Terms In Generalized Gravity,''
  Nucl.\ Phys.\ B {\bf 323}, 242 (1989).

  \bibitem{Hawking-Luttrell} S.~W.~Hawking and J.~C.~Luttrell,
  %``Higher Derivatives In Quantum Cosmology. 1. The Isotropic Case,''
  Nucl.\ Phys.\ B {\bf 247}, 250 (1984).

  \bibitem{Whitt} B.~Whitt,
  %``Fourth Order Gravity as General Relativity Plus Matter,''
  Phys.\ Lett.\ B {\bf 145}, 176 (1984).

  \bibitem{Frolov}  A.~V.~Frolov,
  %``A Singularity Problem with f(R) Dark Energy,''
  Phys.\ Rev.\ Lett.\  {\bf 101}, 061103 (2008)
  [arXiv:0803.2500 [astro-ph]].

  \bibitem{Dyer-Hinterbichler} E.~Dyer and K.~Hinterbichler,
  %``Boundary Terms, Variational Principles and Higher Derivative Modified Gravity,''
  Phys.\ Rev.\ D {\bf 79}, 024028 (2009)
  [arXiv:0809.4033 [gr-qc]].

  \bibitem{Olmo} G.~J.~Olmo,
  %``Limit to general relativity in f(R) theories of gravity,''
  Phys.\ Rev.\ D {\bf 75}, 023511 (2007)
  [gr-qc/0612047].

  \bibitem{Myers} R.~C.~Myers,
  %``Higher Derivative Gravity, Surface Terms And String Theory,''
  Phys.\ Rev.\ D {\bf 36}, 392 (1987).

 \bibitem{Fukuma}  M.~Fukuma, S.~Matsuura and T.~Sakai,
  %``Higher derivative gravity and the AdS / CFT correspondence,''
  Prog.\ Theor.\ Phys.\  {\bf 105}, 1017 (2001)  [hep-th/0103187].

  \bibitem{Cremonini}S.~Cremonini, J.~T.~Liu and P.~Szepietowski,
  %``Higher Derivative Corrections to R-charged Black Holes: Boundary Counterterms and the Mass-Charge Relation,''
  JHEP {\bf 1003}, 042 (2010)  [arXiv:0910.5159 [hep-th]].

 \bibitem{Nojiri}S.~'i.~Nojiri and S.~D.~Odintsov,
  %``Finite gravitational action for higher derivative and stringy gravities,''
  Phys.\ Rev.\ D {\bf 62}, 064018 (2000)  [hep-th/9911152].

  \bibitem{Serg}
  S.~'i.~Nojiri and S.~D.~Odintsov,
  %``On the conformal anomaly from higher derivative gravity in AdS / CFT correspondence,''
  Int.\ J.\ Mod.\ Phys.\ A {\bf 15}, 413 (2000)
  [hep-th/9903033].

  \bibitem{Hohm-Tonni} O.~Hohm and E.~Tonni,
  %``A boundary stress tensor for higher-derivative gravity in AdS and Lifshitz backgrounds,''
  JHEP {\bf 1004}, 093 (2010)
  [arXiv:1001.3598 [hep-th]].

  \bibitem{Woodard}  R.~P.~Woodard,
  %``Avoiding dark energy with 1/r modifications of gravity,''
  Lect.\ Notes Phys.\  {\bf 720}, 403 (2007)
  [astro-ph/0601672].


  \bibitem{Waldram}A.~Hindawi, B.~A.~Ovrut and D.~Waldram,
  %``Nontrivial vacua in higher derivative gravitation,''
  Phys.\ Rev.\ D {\bf 53}, 5597 (1996)
  [hep-th/9509147].

  \bibitem{Imbimbo}C.~Imbimbo, A.~Schwimmer, S.~Theisen and S.~Yankielowicz,
  %``Diffeomorphisms and holographic anomalies,''
  Class.\ Quant.\ Grav.\  {\bf 17}, 1129 (2000)
  [hep-th/9910267].

  \bibitem{Kraus-Larsen}P.~Kraus and F.~Larsen,
  %``Microscopic black hole entropy in theories with higher derivatives,''
  JHEP {\bf 0509}, 034 (2005)
  [hep-th/0506176].


   \bibitem{Saida-Soda}H.~Saida and J.~Soda,
  %``Statistical entropy of BTZ black hole in higher curvature gravity,''
  Phys.\ Lett.\ B {\bf 471}, 358 (2000)
  [gr-qc/9909061].



  \bibitem{Fefferman-Graham-paper}
  C.~Fefferman and C.~R.~Graham,
  {\em Conformal Invariants},
  1985 in {\it Elie Cartan et les Math\'{e}matiques d'aujourd'hui},
  (Ast\'{e}risque vol H S) (Paris: Soc. Math. France) p 95.
  
  \bibitem{David} D.~R.~Karakhanian, R.~P.~Manvelyan and R.~L.~Mkrtchian,
  %``Area preserving structure of 2-d gravity,''
  Phys.\ Lett.\ B {\bf 329}, 185 (1994)
  [hep-th/9401031].



  \bibitem{Schwimmer-Theisen}
  A.~Schwimmer and S.~Theisen,
  %``Entanglement Entropy, Trace Anomalies and Holography,''
  Nucl.\ Phys.\ B {\bf 801}, 1 (2008)
  [arXiv:0802.1017 [hep-th]].

 \bibitem{Waldbook} R.~M.~Wald,
  %``General Relativity,''
  Chicago, Usa: Univ. Pr. (1984) 491p, ISBN 0-226-87033-2.

 \bibitem{Balcerzak} A.~Balcerzak and M.~P.~Dabrowski,
  %``Gibbons-Hawking Boundary Terms and Junction Conditions for Higher-Order Brane Gravity Models,''
  JCAP {\bf 0901}, 018 (2009)
  [arXiv:0804.0855 [hep-th]].

  \bibitem{Cunliff}
  C.~Cunliff,
  %``Non-Fefferman-Graham asymptotics and holographic renormalization in New Massive Gravity,''
  JHEP {\bf 1304}, 141 (2013)
  [arXiv:1301.1347 [hep-th]].



  \bibitem{Sol-BH} S.~N.~Solodukhin,
  %``Holography with gravitational Chern-Simons,''
  Phys.\ Rev.\ D {\bf 74}, 024015 (2006)
  [hep-th/0509148].



 % \bibitem{Wald-paper}V.~Iyer and R.~M.~Wald,  %``Some properties of Noether charge and a proposal for dynamical black hole entropy,''  Phys.\ Rev.\ D {\bf 50}, 846 (1994)  [gr-qc/9403028].

 \bibitem{Tekin} I.~Gullu, T.~C.~Sisman and B.~Tekin,
  %``All Bulk and Boundary Unitary Cubic Curvature Theories in Three Dimensions,''
  Phys.\ Rev.\ D {\bf 83}, 024033 (2011)
  [arXiv:1011.2419 [hep-th]].


 \bibitem{Sotiriou} T.~P.~Sotiriou and V.~Faraoni,
  %``f(R) Theories Of Gravity,''
  Rev.\ Mod.\ Phys.\  {\bf 82}, 451 (2010)
  [arXiv:0805.1726 [gr-qc]].



  \bibitem{vDVZ}
  H.~van Dam and M.~J.~G.~Veltman,
  %``Massive and massless Yang-Mills and gravitational fields,''
  Nucl.\ Phys.\ B {\bf 22}, 397 (1970); V.~I.~Zakharov,
  %``Linearized gravitation theory and the graviton mass,''
  JETP Lett.\  {\bf 12}, 312 (1970)
  [Pisma Zh.\ Eksp.\ Teor.\ Fiz.\  {\bf 12}, 447 (1970)].

  \bibitem{Myung} Y.~S.~Myung,
  %``Graviton and scalar propagations on AdS(4) space in f(R) gravities,''
  Eur.\ Phys.\ J.\ C {\bf 71}, 1550 (2011)
  [arXiv:1012.2153 [gr-qc]].

  \end{thebibliography}
  \end{document}